\documentclass[graybox]{svmult}

\usepackage{mathptmx}       
\usepackage{helvet}         
\usepackage{courier}        
\usepackage{type1cm}        
\usepackage{makeidx}         
\usepackage{graphicx}        
\usepackage{multicol}        
\usepackage[bottom]{footmisc}

\makeindex             

\begin{document}

\title*{Triggering, suppressing and redistributing star formation}
\author{James E. Dale, Barbara Ercolano, Ian Bonnell}
\institute{James E. Dale \at Excellence Cluster `Universe', Boltzmannstrasse 2, 85748, Garching, Germany \email{dale@usm.lmu.de}
\and Barbara Ercolano \at Excellence Cluster `Universe', Boltzmannstrasse 2, 85748, Garching, Germany \email{ercolano@usm.lmu.de}
\and Ian Bonnell \at School of Physics and Astronomy, University of St Andrews, North Haugh, St Andrews, Fife, KY16 9SS, Scotland \email{iab1@st-andrews.ac.uk}}
%
%
\maketitle

\abstract{We discuss three different ways in which stellar feedback may alter the outcome of star cluster formation: triggering or suppressing star formation, and redistributing the stellar population in space. We use detailed Smoothed Particle Hydrodynamics (SPH) simulations of HII regions in turbulent molecular clouds to show that all three of these may happen in the same system, making inferences about the effects of feedback problematic.}

\section{Introduction}
\label{sec:1}
To what degree star formation is self--regulating is much debated in astrophysics. Stellar feedback in the form of HII regions, winds, jets, radiation pressure, non--ionizing radiation and supernova explosions are all potentially able to influence the star formation process in molecular clouds (e.g. \cite{matzner2002}). These processes may have a positive or negative effect on star formation, but positive effects (in the sense of triggering) have received the most observational attention (e.g. \cite{koenig2008}, \cite{puga2009}, \cite{zavagno2010}). However, disentangling how these various feedback mechanisms influence star formation is fraught with difficulty, since it is necessary to think comparatively and infer how star formation would proceed differently if feedback were absent.\\
\indent From this perspective, there are three ways in which feedback may influence the formation of stars. Its effect may be positive (commonly referred to as `triggered star formation'), in the sense of increasing the star formation rate or efficiency, producing more stars, or leading to the birth of stars which would otherwise not exist (note that these effects are not necessarily equivalent and, in the same system, some may transpire while others do not). Feedback may also be negative and do the opposite of these things, which we will call `suppressed star formation'. Of course, the global influence of feedback on a given system may be different from its local effects -- it is perfectly possible for star formation to be suppressed at some locations and triggered in others. Finally, suppression and triggering may cancel each other out and feedback may result in the production of a statistically indistinguishable population of stars, but distribute them differently in position or velocity space relative to their distribution in the absence of feedback, for example in well--defined shells. This could be termed `redistributed star formation'.\\
\indent In order to infer in a given system which of these processes is at work and what is the overall influence on the end product -- the stellar cluster -- it is essential to have a credible counterfactual model for comparison. This is equally true of observed and simulated systems. A good idea of what the system would have done in the absence of feedback can then be gained, and hence the effects of feedback isolated.\\

\section{Numerical simulations}
\label{sec:2}
We have embarked on an SPH parameter--space study in the mass--radius plane of GMCs (fully described in \cite{deb2012}) in which we simulate the effects of the HII regions driven by the massive stars. Control simulations without feedback enable us to study triggering, suppression and redistribution with the benefit of well--defined counterfactual models. Structures such as bubbles, pillars and champagne flows, all of which are commonly associated with feedback and triggered star formation, emerge quite naturally from these calculations. We study the effects of feedback in terms of global parameters, such as the star formation rate and efficiency, and also on a local star--by--star level by inquiring whether the material from which a given star forms is also involved in star formation in the companion run.\\

\section{Results}
\label{sec:3}
We find that the global and local influence of feedback can be very different. In all our simulations, the overall effect of the expanding HII regions on the star formation rates and efficiencies is negative (or negligible). Clouds with escape velocities comparable to the ionized sound speed are largely unaffected. In clouds with low escape velocities, the dense star forming gas near the ionizing stars is rapidly dispersed by feedback; this has the strongest influence on star formation. Many HII regions eventually burst out of the cold gas, becoming champagne flows. The escape from the cloud of the hot HII gas lessens the dynamical effect of feedback. However, using star--by--star comparison with feedback--free control runs, we establish unequivocally that the formation of some stars is triggered, in the sense that the material from which they form is not involved in star formation in the control run. Star formation is thus globally suppressed but sometimes locally triggered.\\
\begin{figure}[t]
\sidecaption
\includegraphics[scale=.35]{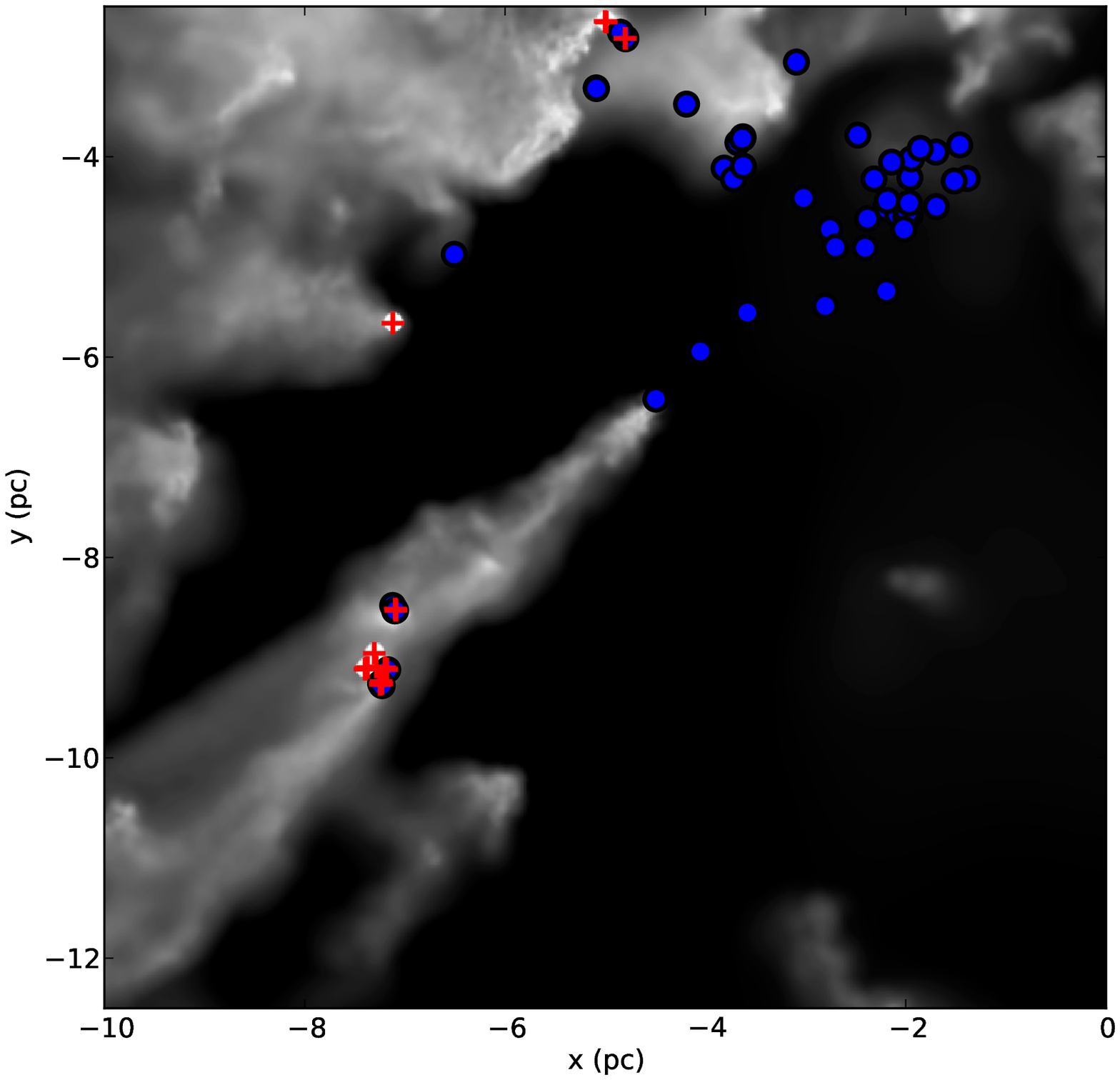}
\hspace{0.01in}
\includegraphics[scale=.35]{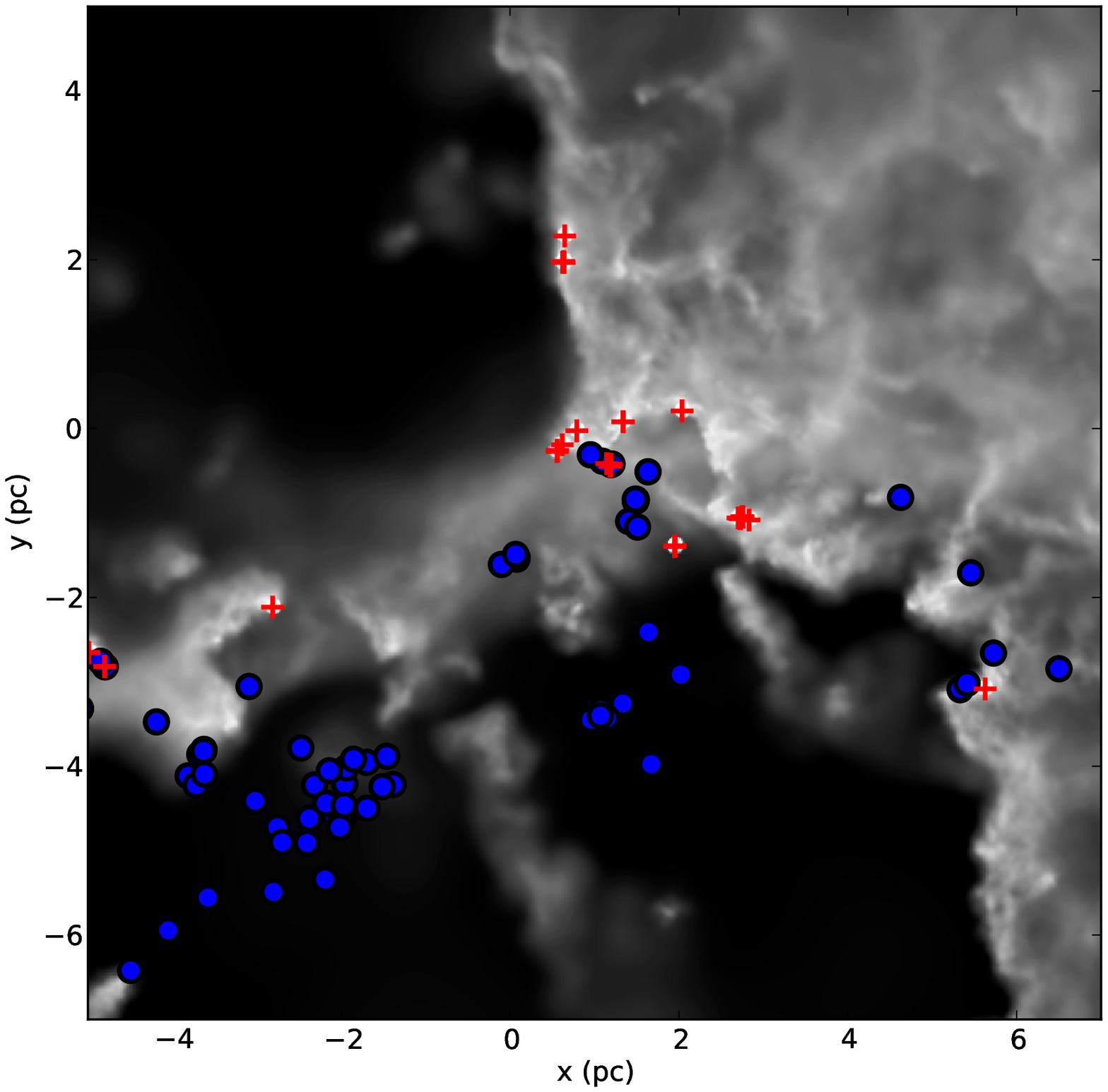}
%
%
\caption{Gas column--density map viewed along the z--axis (greyscale) of a pillar (left panel) and part of a bubble wall (right panel) with red crosses marking the locations of triggered stars and blue circles marking the locations of spontaneously--formed stars.}
\label{fig:1}       
\end{figure}
\indent Triggering is often identified in the literature by the association of young stars with pillars or bubble walls (e.g. \cite{smith2005}, \cite{zavagno2010}, \cite{thompson2012}). In Figure \ref{fig:1} we illustrate the pitfalls of this approach. We show gas column density maps of two regions of the same simulation (Run I from \cite{deb2012}) with overplotted symbols representing triggered (red crosses) and spontaneously--formed (blue circles) stars. Here we define a star to be triggered if less than half the material from which it forms is also involved in star formation in the control simulation. Otherwise, we regard the star as `spontaneous' (see \cite{db2012} for more details on this technique). In the left panel, we show a pillar structure pointing towards the ionizing cluster in this simulation (the group of blue stars in the top right of the image). There are indeed triggered objects associated with the pillar about halfway along its length, but these are mixed with some spontaneously--formed stars and those objects nearest the pillar tip are also in fact formed spontaneously. The pillar in the ionized run is the remains of an accretion flow and the gas from which the stars nearest the tip form is, in the control run, simply delivered to the central cluster, where it is involved in star formation.\\
\indent Similarly, in the right panel, we show part of the wall of the bubble structure excavated by the HII regions, which contains a mixture of triggered and spontaneously--formed stars. This admixture is a consequence of the sweeping up, into the same location, of material which was going to form stars anyway, and of quiescent gas which was not. This is an indication that even those stars which form spontaneously may be found in very different positions in the ionized run compared with the control run, an example of redistributed star formation.\\

\section{Summary}
We consider three ways in which feedback influences star cluster formation: triggering and suppression which, respectively, positively or negatively alter the rate or efficiency of star formation and/or the formation of individual stars, and redistribution, which alters the geometrical distribution of stars, even if other properties of the stellar population are unchanged. All these effects may be global or local.\\
\indent We find in our SPH simulations of HII regions driven into turbulent clouds that all three outcomes are present. Although the global effect of feedback is to decrease the star formation rate and efficiency and the formation of many stars is aborted, local triggering does occur. In addition, the stars which we know from our control simulations form spontaneously are to be found in very different locations in the ionized simulations, due to the general sweeping up and transport of star--forming gas by the HII regions. This is a clear example of redistributed star formation. We note therefore that the association of young stars with structures such as pillars or bubble walls is not necessarily a reliable indicator of triggering.\\
\indent We stress the need for credible counterfactuals model when discussing the purported effects of feedback. In order to make reliable statements about what effect feedback has had on a given system, it must be possible to infer at least approximately what the properties of the system would have been in its absence, and to define very carefully in what ways the real system is different.

\end{document}